\documentclass[prl,amsmath,amssymb,superscriptaddress,twocolumn]{revtex4-1}
\usepackage{graphics}
\usepackage{graphicx}
\usepackage{bm}

\newcommand{\dd}{{\bm \delta}}

\newcommand{\mat}[4]{\left(\begin{smallmatrix}#1 & #2\\#3 & #4\end{smallmatrix}\right)}
\DeclareMathAlphabet{\mathpzc}{OT1}{pzc}{m}{it} 
\begin{document}
\title{Universal probes of two-dimensional topological insulators: Dislocation and $\pi$-flux}
\author{Vladimir Juri\v ci\' c}
\affiliation{Instituut-Lorentz for Theoretical Physics, Universiteit
Leiden,
 P.O. Box 9506, 2300 RA Leiden, The Netherlands}
\author{Andrej Mesaros}
\affiliation{Department of Physics, Boston College, Chestnut Hill, MA 02467, USA}
\affiliation{Instituut-Lorentz for Theoretical Physics, Universiteit
Leiden,
 P.O. Box 9506, 2300 RA Leiden, The Netherlands}
\author{Robert-Jan Slager}
\affiliation{Instituut-Lorentz for Theoretical Physics, Universiteit
Leiden,
 P.O. Box 9506, 2300 RA Leiden, The Netherlands}
\author{Jan Zaanen}
\affiliation{Instituut-Lorentz for Theoretical Physics, Universiteit
Leiden,
 P.O. Box 9506, 2300 RA Leiden, The Netherlands}
\begin{abstract}
We show that the $\pi$-flux and the dislocation represent topological observables that probe two-dimensional topological order through binding of the zero-energy modes. We analytically demonstrate that $\pi$-flux hosts a Kramers pair of zero modes in the topological $\Gamma$ (Berry phase skyrmion at the zero momentum) and $M$ (Berry phase skyrmion at a finite momentum) phases of the $M-B$ model introduced for the HgTe quantum spin Hall insulator. Furthermore, we analytically show that the dislocation acts as a $\pi$-flux, but only so in the $M$ phase. Our numerical analysis confirms this through a Kramers pair of zero modes bound to a dislocation appearing in the $M$ phase only, and further demonstrates the robustness of the modes to disorder and the Rashba coupling.
Finally, we conjecture that by studying the zero modes bound to dislocations all translationally distinguishable two-dimensional topological band insulators can be classified.
\end{abstract} 

\maketitle

The topological band insulators (TBIs) in two and three dimensions have attracted a great interest in both theoretical and experimental condensed matter physics\cite{hasan-review,qi-zhang-review}. An extensive classification of topological phases in free fermion systems with a bulk gap, based on time-reversal symmetry (TRS) and particle-hole symmetry (PHS), has been provided \cite{classification}, and it defines the so-called tenfold way. 
This however relies on the spatial continuum limit while TBIs actually need a crystal lattice which, in turn, breaks the translational symmetry. Therefore, the question arises whether the crystal lattice may give rise to an additional subclass of topological phases.  The simple $M-B$ model introduced for the two-dimensional (2D) HgTe quantum spin Hall insulator\cite{bernevig} gives away a generic wisdom in this regard.  Depending on its parameters, this model describes topological phases which are in a thermodynamic sense distinguishable: their topological nature is characterized by a Berry phase skyrmion lattice (SL) in the extended Brillouin zone (BZ), where the sites of this lattice coincide with the reciprocal lattice vectors (``$\Gamma$-phase") or with the TRS points $(\pi,\pi)$ (``M-phase").

The question arises how to distiguish these phases by a topological observable. Early on it was observed in numerical calculations that a TRS localized magnetic $\pi$ flux binds zero-modes in the $\Gamma$-phase\cite{dung-hai2,dung-hai1,qi-zhang}. It was also discovered that screw dislocation lines in three-dimensional TBIs carry zero mode structure\cite{ying-vishwanath}. Here we will demonstrate the special status of such topological defects as the universal bulk probes of the electronic topological order. We will present an analytical description of the zero modes bound to $\pi$ flux, demonstrating that these are closely related, but yet different from the well-known Jackiw-Rossi solutions\cite{jackiw-rossi}. These modes also generalize to two-dimensions the effect of one-dimensional spin-charge separation\cite{SSH}.  By performing numerical computations, we will show that the dislocation binds zero modes only in the topologically nontrivial $M$-phase. Subsequently, we will demonstrate that, modulo a change of basis, these dislocation zero modes are of the same kind as the $\pi$ flux ones, but appear only in the case of non-$\Gamma$-type TBIs. We conjecture that by studying the presence of zero modes associated with dislocations all possible ``translationally distinguishable" TBIs can be classified at least in the 2D case.

Consider a tight-binding model proposed to describe HgTe quantum wells \cite{bernevig}
\begin{equation}\label{eq:tight-binding}
{\mathcal H}=\sum_{{\bf k}}\Psi^\dagger({\bf k})\left(\begin{array}{cc}H({\bf k})& 0\\
0 & H({\bf k})\end{array}\right)\Psi({\bf k})
\end{equation}
where $\Psi^\top=(u_\uparrow,v_\uparrow,u_\downarrow,v_\downarrow)\equiv(\Psi_\uparrow,\Psi_\downarrow)$, while $u$ and $v$ represent two low-energy orbitals, $E1$ and $H1$ in the case of the HgTe system. The upper and the lower blocks in the Hamiltonian are related by the time-reversal symmetry, and $H({\bf k})$ acting in the orbital space has the form
\begin{equation}\label{eq:upper-ham}
H({\bf k})=\tau_\mu d_\mu({\bf k}),
\end{equation}
where
$\tau_\mu$, $\mu=1,2,3$, are the Pauli matrices, $d_{1,2}({\bf k})=\sin k_{x,y}$, and $d_3=M-2B(2-\cos k_x-\cos k_y)$, while lattice constant $a=1$, $\hbar=c=e=1$, and summation over the repeated indices are assumed hereafter. The Dirac mass $M$ and the Schr\" odinger mass $B$
are  related to the material parameters, while the form of the Hamiltonian (\ref{eq:tight-binding}) is set by the symmetries of the system\cite{konig}.
Since the above Hamiltonian has the spectrum $E({\bf k})=\sqrt{d_\mu d_\mu}$ doubly degenerate in the spin space, the band gap opens at the $\Gamma$-point  for  $0<M/B<4$ and the system is in a topologically nontrivial $\Gamma$-phase. For $4<M/B<8$, the system is still topologically nontrivial, but with the bandgap opening at ${\bf k}=(\pi,0)$ and $(0,\pi)$
($M$-phase). On  the other hand, for $M/B<0$ and $M/B>8$ the system is topologically trivial with the bandgap located at the $\Gamma$ and the $M$ point, respectively.

\begin{figure*}
    \includegraphics[width=\textwidth]{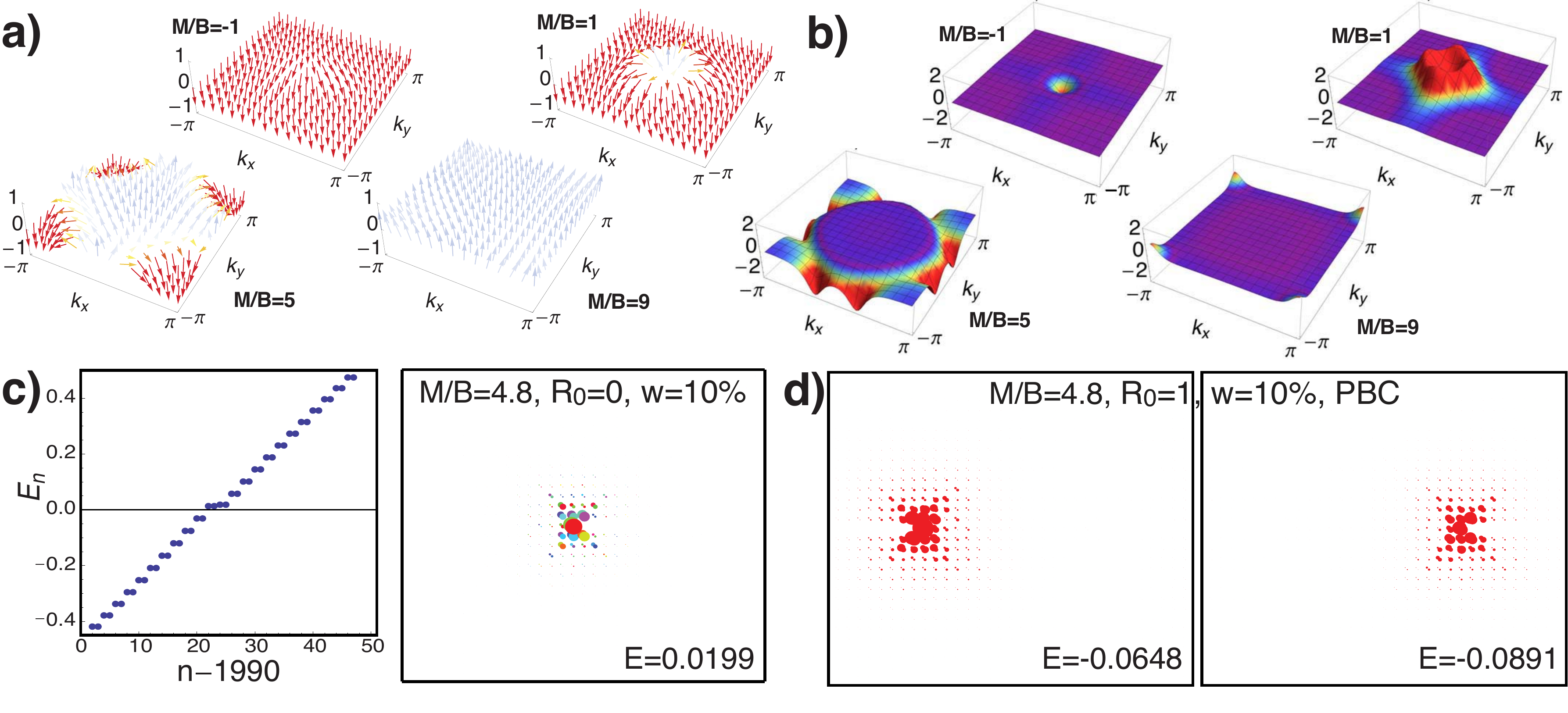}
    \caption{\label{fig:1} (color online) Model of Eq.~(\ref{eq:tight-binding}) in the BZ: (a) Band-structure $\hat{d}(\bf k)$. (b) Skyrmion density $s(\bf k)$. Mid-bulkgap localized dislocation states in $33\textrm{x}30$ unit-cell $M-B$ tight-binding lattice with disorder. The Kramers degenerate pair states are omitted. (c) Dislocation in the center. Offset disks represent the amplitude of $s\uparrow,p\uparrow$ states, and the color their phase. (d) Total wavefunction amplitude in a periodic system (necessitating two dislocations), with Rashba coupling ($R_0$) mixing spins. }
\end{figure*}

In the topologically nontrivial $\Gamma$-phase, the band-structure vector field $\hat{{\bf d}}({\bf k})\equiv {\bf d}({\bf k})/|{\bf d}({\bf k})|$ [Eq.~\eqref{eq:upper-ham}] forms a skyrmion centered at the $\Gamma$-point in the BZ, Fig.\ \ref{fig:1}(a) for $M/B=1$, with the corresponding skyrmion density $s({\bf k})\equiv \hat{{\bf d}}({\bf k})\cdot(\partial_{k_x}\hat{{\bf d}}({\bf k})\times\partial_{k_y}\hat{{\bf d}}({\bf k}))$ shown in Fig.\  \ref{fig:1}(b) [$s({\bf k})$ tracks the position of minimal bandgap in the BZ, coinciding with it where ${\hat{\bf d}}({\bf k}) || {\partial_{ k_x}} {\hat{\bf d}}({\bf k})\times  {\partial_{ k_y}} {\hat{\bf d}}({\bf k})$].  In the 2D extended BZ, this skyrmion structure forms a lattice which respects point group symmetry of the original square lattice. Furthermore, in the $M$-phase, the skyrmion is centered at the $M$ point in the BZ, Fig.\ \ref{fig:1}(a) for $M/B=5$. The position of the corresponding
skyrmion lattice relative to the extended BZ is therefore different than in the $\Gamma$-phase, but the skyrmion lattice still respects the point group symmetry of the square lattice. On the other hand, in the topologically trivial phase, the vector field ${\bf d}$ forms no skyrmion in the BZ, as shown in Fig.\  \ref{fig:1}(a) for $M/B=-1$ and $M/B=9$, consistent with the vanishing of topologically invariant spin Hall conductance $\sigma^S_{xy}=(4\pi)^{-1}\int_{BZ}d^2{\bf k}\;s({\bf k})$. The position of the skyrmion lattice relative to the extended BZ thus encodes translationally active topological order which, as we show below, is probed by the lattice dislocations.

Let us first numerically demonstrate that lattice dislocations bind zero modes as robust topological phenomena in the $M$-phase of the $M-B$ model. Analysis in Ref.~\cite{ying} predicts that PHS (which our model respects), topologically protects localized zero modes on dislocations when the system has $K_{edge}=\pi$ edge modes which our model has only in the $M$ phase.  We indeed find in the $M$-phase a Kramers pair of zero modes bound to a dislocation which results from the fact that, as we show below, the dislocation acts as a $\pi$-flux in this phase. However, we here show the robustness of these modes in the bulk gap when we introduce the random chemical potential, and thus break PHS. To this end we performed numerical analysis of the tight-binding $M-B$ model in the real space
\begin{eqnarray}
  \label{eq:1}
  H_{TB}&=&\sum_{{\bf R},\dd}\Bigg(\Psi_{\bf R}^\dagger\left[ T_\dd+i\frac{R_0}{2}(\openone+\tau_3){\bf e}_z\cdot({\bm \sigma}\times\dd)\right]\Psi_{{\bf R}+\dd}\nonumber\\
&+&\Psi_{\bf R}^\dagger\frac{\epsilon}{2}\Psi_{\bf R}+H.c.\Bigg),
\end{eqnarray}
where $\Psi_{\bf R}=(s_\uparrow({\bf R}),p_\uparrow({\bf R}),s_\downarrow({\bf R}),p^*_\downarrow({\bf R}))$ annihilates the $|E1,\frac{1}{2}\rangle\sim|s\rangle$ type, and $|H1,\frac{3}{2}\rangle\sim|p_x+i p_y\rangle$ type orbitals at site ${\bf R}$ and nearest neighbors $\dd\in\{{\bf e}_x,{\bf e}_y\}$; Pauli matrices ${\bm \sigma}$ mix spin. We set $T_{\dd,\uparrow\uparrow}=\mat{\Delta_s}{t_\dd/2}{t'_\dd/2}{\Delta_p}$, $T_{\dd,\downarrow\downarrow}=T^*_{\dd,\uparrow\uparrow}$, with $t_x=t'_x=-i$, $t_y=-t'_y=-1$, $\Delta_{s/p}=\pm B+D$ and on-site energies $\epsilon=[(C-4D)\tau_0+(M-4B)\tau_3]\otimes\sigma_0$. This reproduces Eq.~\eqref{eq:upper-ham} when $C=D=0$, implying vanishing chemical potential. The $R_0$ term is the nearest neighbor Rashba spin-orbit coupling \cite{rothe} which is induced by broken $z\rightarrow -z$  reflection symmetry of the quantum well, and should be relevant in case of tunneling measurements on thin wells.

Our numerical analysis of $H_{TB}$ pertains to various system shapes and sizes, with varying disorder strengths given by multiplication of the parameters for each ${\bf R},\dd$ by Gaussian random variables of width $w$, while preserving TRS.  Fig.~\ref{fig:1}(c) demonstrates the spectrum and wavefunction at $w=10\%$ disorder. Localization (even by weak disorder) decouples dislocation states from edges and possible edge roughness effects.

\begin{figure*}
  \includegraphics[width=\textwidth]{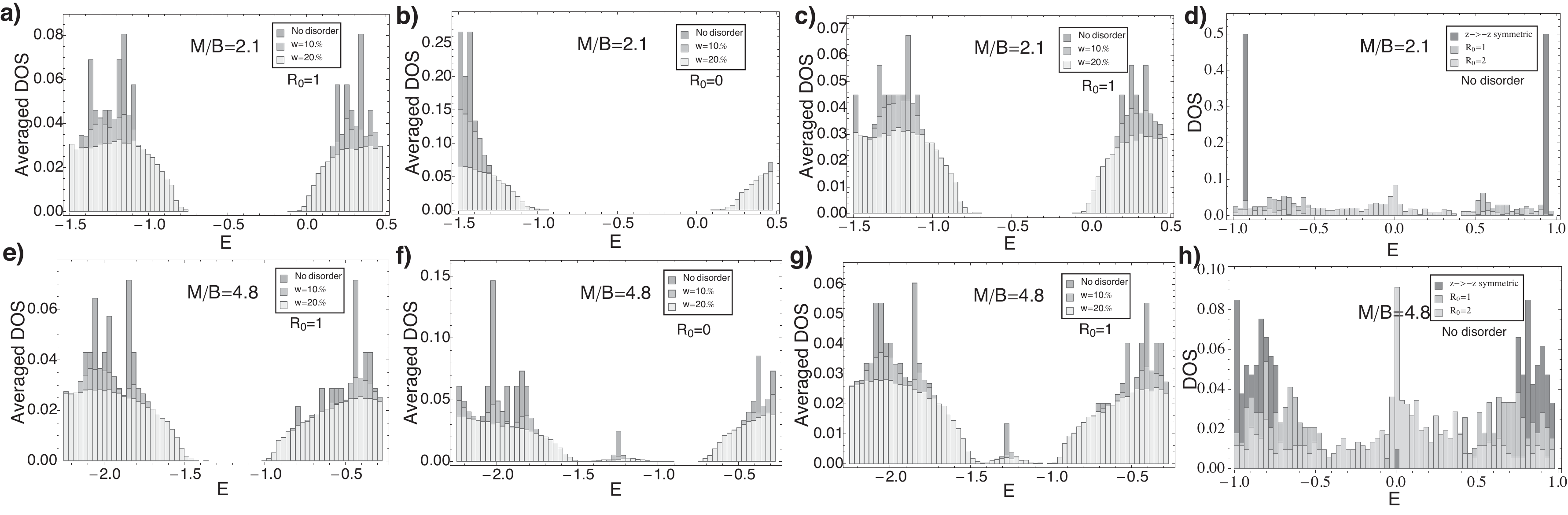}
\caption{\label{fig:2} Comparison of $\Gamma$ (a-d) and $M$ (e-h) phases. The density of states of $21\textrm{x}18$ lattice (100 disorder realizations averages), with $C\equiv 0.2|t|$, $D\equiv 0.3|t|$ setting the chemical potential, and $R_0$ the Rashba coupling. (a) and (e) In absence of dislocation. (b,f) Robust midgap dislocation modes are present only in $M$-phase;  (c,g) Same is true upon spin mixing through $R_0$. (d,h) Strong Rashba coupling closes the topological bulkgap.}
\end{figure*}

In the presence of the Rashba coupling ($R_0\neq 0$, but not large enough to close the topological bulk gap~\cite{kane-mele1}), the spins are mixed, but the Kramers pairs remain localized (Fig.~\ref{fig:1}d). Fig.~\ref{fig:2} demonstrates the robustness of dislocation modes within the topological bulkgap, through the Rashba coupling perturbed and disorder averaged density of states (DOS) of $H_{TB}$ in a periodic lattice, contrasted between the $\Gamma$ and the $M$ phase.

Let us now analytically show, using the elastic continuum theory, that the effect of a lattice dislocation in the $M$-phase is to effectively introduce a magnetic $\pi$ flux. We consider a dislocation with Burgers vector ${\bf b}$, and expand the Hamiltonian (\ref{eq:upper-ham}) around the M-point in the BZ. As a next step, a dislocation introduces an elastic deformation of the medium described by the distortion $\{{\bm \varepsilon}_i({\bf r})\}$ of the (global) Cartesian basis $\{{\bf e}_i\}$, $i=x,y$, in the tangent space at the point ${\bf r}$\cite{LL-elasticity}, the momentum in the vicinity of the $M$-point reads 
\begin{equation}
k_i={\bf E}_i\cdot({\bf k}_M-{\bf q})=({\bf e}_i+{\bm \varepsilon}_i)\cdot({\bf k}_M-{\bf q}),
\end{equation} 
where ${\bf k}_M=(\pi,\pi)/a$, ${\bf q}$ is the momentum of the low-energy excitations, $|{\bf q}|\ll |{\bf k}_M|$, and we have restored the lattice constant $a$. 
The corresponding continuum Hamiltonian after this coarse graining\cite{BST} is 
\begin{equation}\label{eq:cont-ham-M}
H_{\rm eff}({\bf k},{\bf A})=\tau_i(k_i+A_i)+[{\tilde M}-{\tilde B}({\bf k}+{\bf A})^2]\tau_3,
\end{equation}
with redefinition ${\bf q}\rightarrow{\bf k}$, ${\tilde M}\equiv M-8B$, ${\tilde B}\equiv-B$, and $A_i\equiv-{\bm \varepsilon}_i\cdot{\bf k}_M$. The form of the distortion ${\bm \varepsilon}_i$ is determined from the dual basis in the tangent space at the point ${\bf r}$ which in the case of a dislocation with ${\bf b}=a {\bf e}_x$ is ${\bf E}^x=(1-\frac{ay}{2\pi r^2} ){\bf e}^x+\frac{ax}{2\pi r^2}{\bf e}^y, {\bf E}^y={\bf e}^y$\cite{kleinert}. Using that ${\bf E}_i\cdot{\bf E}^j=\delta_i^j$, we obtain the distortion field to the leading order in $a/r$,  ${\bm\varepsilon}_x=\frac{ay}{2\pi r^2}{\bf e}_x,{\bm\varepsilon}_y=-\frac{ax}{2\pi r^2}{\bf e}_y $.
Finally, this form of the distortion yields the vector potential 
\begin{equation}\label{eq:pi-flux}
{\bf A}=\frac{-y{\bf e}_x+x{\bf e}_y}{2r^2}
\end{equation}
in Eq.\ (\ref{eq:cont-ham-M}), and therefore the dislocation in the $M$-phase acts as a magnetic $\pi$-flux. On the other hand, in the $\Gamma$-phase, the continuum Hamiltonian has the generic form (\ref{eq:cont-ham-M}), with ${\tilde M}=M$ and ${\tilde B}=B$, but the action of the dislocation is trivial, since the bandgap is at zero momentum, and thus ${\bf A}=0$.

The last operation is to show the origin of the zero-modes bound to the $\pi$-flux/dislocation. This involves a generalization of the Jackiw-Rossi (JR) mechanism\cite{jackiw-rossi}. We find that instead of the vortex in the mass required for the JR solutions, quite similar zero modes are formed by fermions with a momentum-dependent mass in the background of a $\pi$-flux. 
In the $\Gamma$ phase, we express the Hamiltonian (\ref{eq:cont-ham-M})  in polar coordinates $(r,\varphi)$, with ${\tilde M}=M$, ${\tilde B}=B$, the vector potential as in Eq.\ (\ref{eq:pi-flux}),
and use the ansatz for the spin up zero-energy state
\begin{equation}\label{eq:zero-energy-spinor}
\Psi(r,\varphi)=\left(\begin{array}{cc}e^{i(l-1)\varphi}u_{l-1}(r)\\
e^{il\varphi}v_l(r)\end{array} \right),
\end{equation}
with $l\in \mathbb{Z}$ as the angular momentum quantum number and the spin index for the spinor suppressed. The function $u$ is then found to be solution of the following equation:
\begin{equation}\label{eq:zero-mode-u}
\left[M^2+(2MB-1){\cal O}_{l-\frac{1}{2}}+B^2{\cal O}_{l-\frac{1}{2}}^2\right]u_{l-1}(r)=0,
\end{equation}
where the operator ${\cal O}_l\equiv \partial_r^2+r^{-1}\partial_r-r^{-2}l^2$ .
The function $v_l(r)$  obeys an equation of the same form as (\ref{eq:zero-mode-u}) with $l\rightarrow l+1$.

This result also follows by noting that if the spinor in Eq.\ (\ref{eq:zero-energy-spinor}) is an eigenstate with zero eigenvalue of $H_{\rm eff}({\bf k},{\bf A})$, then it is also an eigenstate with the same eigenvalue of the square of $H_{\rm eff}({\bf k},{\bf A})$. One then obtains
\begin{equation}
H_{\rm eff}({\bf k},{\bf A})^2=B^2 ({\tilde{\bf k}}^2)^2+(1-2MB){\tilde{\bf k}}^2+M^2,
\end{equation}
with ${\tilde{\bf k}}\equiv{\bf k}+{\bf A}$, and the operator ${\tilde{\bf k}}^2$ after acting on the angular part of the upper component of the spinor (\ref{eq:zero-energy-spinor}) yields Eq.\ (\ref{eq:zero-mode-u}).  From Eq.\ (\ref{eq:zero-mode-u}) we conclude that the function $u_{l-1}(r)$ is an eigenfunction of the operator ${\cal O}_{l-1/2}$ with a {\it positive} eigenvalue
\begin{equation}\label{eq:u}
{\cal O}_{l-\frac{1}{2}}u_{l-1}(r)=\lambda^2 u_{l-1}(r),
\end{equation}
since the operator ${\tilde{\bf k}}^2$ when acting on a function with the angular momentum $l$ is equal to $-{\cal O}_{l+1/2}$, and the eigenstates of the  operator ${\tilde{\bf k}}^2$
with a {\it negative} eigenvalue are localized.  Eqs.\ (\ref{eq:zero-mode-u}) and (\ref{eq:u}) then imply
\begin{equation}\label{eq:lambda-square}
\lambda_\pm=\frac{1\pm\sqrt{1-4MB}}{2B},
\end{equation}
and the function $u_l(r)$ is a linear combination of the modified Bessel functions of the first and the second kind, $I_{l-1/2}(\lambda r)$ and $K_{l-1/2}(\lambda r)$, respectively.  Their asymptotic behavior at the origin and at infinity then implies that the only square-integrable solutions are in the zero angular-momentum channel. Note that the localization lengths $\lambda_\pm$ coincide with the penetration depths associated with the edge modes.
We should now distinguish two regimes of parameters, $0<MB<1/4$ and $MB>1/4$, for which the argument of the square-root in Eq.(\ref{eq:lambda-square}) is positive and negative, respectively.

For $0<MB<1/4$, $\lambda_\pm$ are purely real, and using Eq.\ (\ref{eq:zero-mode-u})
we obtain zero-energy solution in the form
\begin{equation}\label{eq:zero-energy-states1}
\Psi({\bf r})=\Psi_+({\bf r})+\Psi_-({\bf r}),
\end{equation}
where
\begin{equation}\label{eq:psi-plus-minus1}
\Psi_\pm({\bf r})\equiv \frac{1}{\sqrt{r}}\left(C_1^{\pm}e^{\mp\lambda_+r}+C_2^{\pm}e^{\mp\lambda_-r}\right)\left(\begin{array}{cc}e^{-i\varphi}\\ \pm i\end{array}\right),
\end{equation}
with $C_{1,2}^\pm$ complex constants.
For both $M$ and $B$ positive (negative),
$\lambda_\pm>0$ ($\lambda_\pm<0$), and therefore only the solution $\Psi_+$ ($\Psi_-$) is normalizable.
For $MB>1/4$, the existence of the zero-modes can be similarly shown.

The zero-energy modes (\ref{eq:psi-plus-minus1})
form an overcomplete basis in the zero angular-momentum channel as a consequence of the singularity of the vortex potential.
A particular regularization provided by considering the vortex with the flux concentrated in a thin annulus ensures Hermitianity of the Hamiltonian~\cite{melikyan,JHT}. This regularization yields, up to normalization,
\begin{equation}\label{eq:zero-energy-annulus}
\Psi({\bf r})=\frac{e^{-\lambda_+r}-e^{-\lambda_-r}}{\sqrt{r}}\left(\begin{array}{cc}e^{-i\varphi}\\ i\end{array}\right).
\end{equation}
Notice that this zero-energy state is regular at the origin which is a consequence of the regularity at the origin of the solutions in the absence of the vortex. Other regularizations, representing different microscopic conditions at the vortex, can lead to different behavior~\cite{us-PRB}.

This result for the bound states to the vortex in the quantum spin Hall state should be compared with the one for a trivial insulator  when $MB<0$, and to be concrete we consider $B<0$. Then, $\lambda_+<0$ and $\lambda_->0$, and normalizable solutions are given by Eqs.\ (\ref{eq:zero-energy-states1}) and (\ref{eq:psi-plus-minus1}) with $C_1^+=C_2^-=0$.
However, since the upper and the lower component of the spinor diverge at the origin with opposite signs, the condition of regularity cannot be satisfied simultaneously for both, and therefore no zero-energy mode exists in the topologically trivial phase. Notice that the condition of regularity at the origin is analogous to the boundary condition  for a TI interfaced with the trivial vacuum\cite{konig}, while the flux provides topological frustration to the electrons in the bulk.

These results also pertain to the $\pi$-flux in the topologically nontrivial $M$-phase  ${\tilde M}/{\tilde B}>0$ in Eq.\ (\ref{eq:cont-ham-M}), now involving the coarse grained states around the $M$ point, where we also find zero-modes bound to $\pi$-flux. Since the dislocation in the $M$-phase, as we have shown, effectively acts as a $\pi$-flux, 
it also binds a pair of zero modes in this phase, and in fact, this explains our numerical resuls. On the other hand, in the trivial phase 
${\tilde M}/{\tilde B}<0$-neither the $\pi$-flux nor the dislocation binds zero modes.

Depending on their occupation, the modes bound to the topological defects carry nontrivial charge or spin quantum number.  The spin-charge separation, characteristic for one-dimensional systems\cite{SSH}, thus appears also in a two-dimensional system, and is tied to a topologically non-trivial nature of the quantum spin Hall state\cite{dung-hai1,qi-zhang}.

These findings are obviously consequential for experiment. Dislocations are ubiquitous in any real crystal, for instance in the form of small angle grain boundaries. We predict that their cores should carry zero modes, which should be easy to detect with scanning tunneling spectroscopy. The experimental challenge just lies in the realization of non-$\Gamma$ TIs that are also easily accessible to spectroscopic measurements.

Based on these specific results we conjecture the following general principle. Besides the ``translationally trivial" $\Gamma$-type TBIs, completely classified by the tenfold way, there is a further subclassification in terms of  ``translationally active" TBIs which are in 2D characterized by the locus of the Berry phase skyrmion lattice in the extended BZ, involving high symmetry points respecting time-reversal invariance and point group symmetries (like the $M$ phase). The $\pi$-fluxes and dislocations act as the topological bulk probes  and the presence or absence of zero modes can be used as a classification tool for the translational side of topological order in TBIs. For instance, the Haldane phase on the honeycomb lattice\cite{haldane-PRL} has skyrmions at the valleys and accordingly zero modes bound to dislocations. At present we are testing this hypothesis for all wallpaper groups in 2D.

The authors thank X.-L.\ Qi, I.\ Herbut, C.\ F.\ J.\ Flipse, and Y.\ Ran for useful discussions. V.\ J.\ acknowledges the support of the Netherlands Organization for Scientific Research (NWO).

\end{document}